\documentclass[preprint,aps]{revtex4}
\usepackage{graphicx}
\usepackage{dcolumn}
\usepackage{bm}

\begin{document}


\title {Hall effect in PrB$_6$ and NdB$_6$}

\author{M.A.Anisimov$^{1, 2}$} \author{A.V.Bogach$^{1}$}\author{V.V.Glushkov$^{1, 2}$}
\author{S.V.Demishev$^{1}$}\author{N.A.Samarin$^{1}$}
\author{V.B.Filipov$^{3}$} \author{N.Yu.Shitsevalova$^{3}$}
\author{N.E.Sluchanko$^{1,}$} \email{nes@lt.gpi.ru}

\affiliation{$\phantom{x}^1$--A.M. Prokhorov General Physics
Institute of RAS\\ 38 Vavilov str., Moscow, 119991 Russia}

 \affiliation{$\phantom{x}^2$--Moscow Institute of Physics and Technology
(State University)\\ 9 Institutskii per., Dolgoprudnyi, Moscow
Region 141700 Russia}

\affiliation{$\phantom{x}^3$--Institute for Problems of Materials
Science of NAS\\ 3 Krzhizhanovskii str., Kiev, 03680 Ukraine}

\date{\today}

\begin{abstract}
Hall effect was studied on the single crystals of antiferromagnets
PrB$_6$ and NdB$_6$ at temperatures $2$K$<$T$<300$K in magnetic
fields up to $8$T using the sample rotation technique. At low
magnetic fields $\mu_0$H$\leq1$T Hall coefficient R$_\mathrm{H}$,
which is practically temperature independent in paramagnetic state
at $8$K$\leq$T$\leq70$K, is characterized by the values of
R$_H$(PrB$_6$)$\sim-(4.2\pm0.1)\cdot10^{-4}$ cm$^3$/C and
R$_H$(NdB$_6$)$\sim-(4.1\pm0.1)\cdot10^{-4}$ cm$^3$/C. Rather
different behaviour of R$_H$ is observed in antiferromagnetic (AF)
phases of these hexaborides. For PrB$_6$ the decrease of
temperature below T$_N\approx6.7$K is accompanied by a noticeable
($\Delta R_H/R_H\sim10\%$) elevation of R$_H$($\mu_0$H$=1$T) to
the values of $-(3.8\pm0.1)\cdot10^{-4}$ cm$^3$/C. On the
contrary, the low field Hall coefficient in NdB$_6$ diminishes by
about $15\%$ reaching the value
R$_H\approx-(4.7\pm0.1)\cdot10^{-4}$ cm$^3$/C in AF state at
$2.5$K. The increase of magnetic field inducing magnetic
transition in the commensurate magnetic phase of PrB$_6$ results
in essential R$_H$ changes (up to $10\%$) at liquid helium
temperatures. The anomalous behaviour of the charge transport
parameters for RB$_6$ (R=Pr, Nd) found in vicinity of Neel
temperature suggests the possible effect of $5d$-states spin
density polarization of both in AF and paramagnetic states of the
compounds under investigation.
\end{abstract}

\pacs{Valid PACS should be here}


\maketitle

\section*{1.INTRODUCTION}

The interest to the family of rare earth hexaborides (RB$_6$) is
supported by their promising applications as thermoelectric and
effective thermionic cathode materials \cite{Sl1, Ta2}. However,
these compounds also demonstrate an exceptional variety of unusual
physical phenomena. In particular, the heavy fermion compound
CeB$_6$ was recently shown to enter into unusual AF phase with
anomalous transport and magnetic properties \cite{Sl3}. SmB$_6$ is
known to be an archetypal intermediate valence compound with fast
charge fluctuations \cite{Wach4}. Europium hexaboride (EuB$_6$)
demonstrates the colossal magnetoresistance effect in the vicinity
of ferromagnetic phase transition \cite{Sul5}. Finally, two
consecutive phase transitions - structural one induced by
cooperative Jahn-Teller effect with changing the symmetry from
cubic to rhombohedral and magnetic one with AF ordering into a
complicated triple-k spin structure are detected in DyB$_6$
\cite{Tak6}.

In terms of the $4f$-shell filling of rare-earth ion
antiferromagnets PrB$_6$ and NdB$_6$ share the places in the
RB$_6$ sequence between heavy fermion system CeB$_6$ and
intermediate valence compound SmB$_6$. In spite of very similar
paramagnetic Fermi surfaces (FS) of LaB$_6$, PrB$_6$ and NdB$_6$
resulting to a common indirect RKKY-exchange motive the magnetic
moments of $4f^2$(Pr) and $4f^3$(Nd) configurations are arranged
into different magnetic structures. In particular, when lowering
the temperature incommensurate AF phase (IC) formed in PrB$_6$
below Neel temperature T$_N\sim7$K evolves to commensurate AF one
(C) observed at T$<$T$_M\sim4.2$K \cite{Ser7}-\cite{Iwa9} (see
Fig.1a). Application of magnetic field changes the magnetic unit
cell of PrB$_6$ inducing the magnetic transition from C phase to
another commensurate C$_H$ phase \cite{Ser7}-\cite{Iwa9} (Fig.1a).
On the contrary, only one commensurate AF structure is detected in
NdB$_6$ at T$<$T$_N\sim8$K in magnetic fields $\mu_0$H$<15$T
\cite{Awa10}-\cite{Yon12} (Fig.1b).

It should be noted here that the Fermi surface of NdB$_6$ in the
AF state differs noticeably from that of PrB$_6$. Indeed, in
addition to the main FS fragment of RB$_6$ formed by large
X-centered ellipsoids connected by the necks in X-X directions
\cite{Onu13, End14} new regions centered at R-points are evidently
established in the AF state of NdB$_6$ from the calculations of
new specific branches detected in quantum oscillation experiments
\cite{Kub15}. These additional folded fragments of FS induced by
AF ordering could modify the strength of indirect exchange between
the $4f$-shell magnetic moments thus resulting in the different
structure and parameters of the magnetic H-T phase diagrams for
PrB$_6$ and NdB$_6$ (Fig.1).

\begin{figure*}
\includegraphics[scale=1.2]{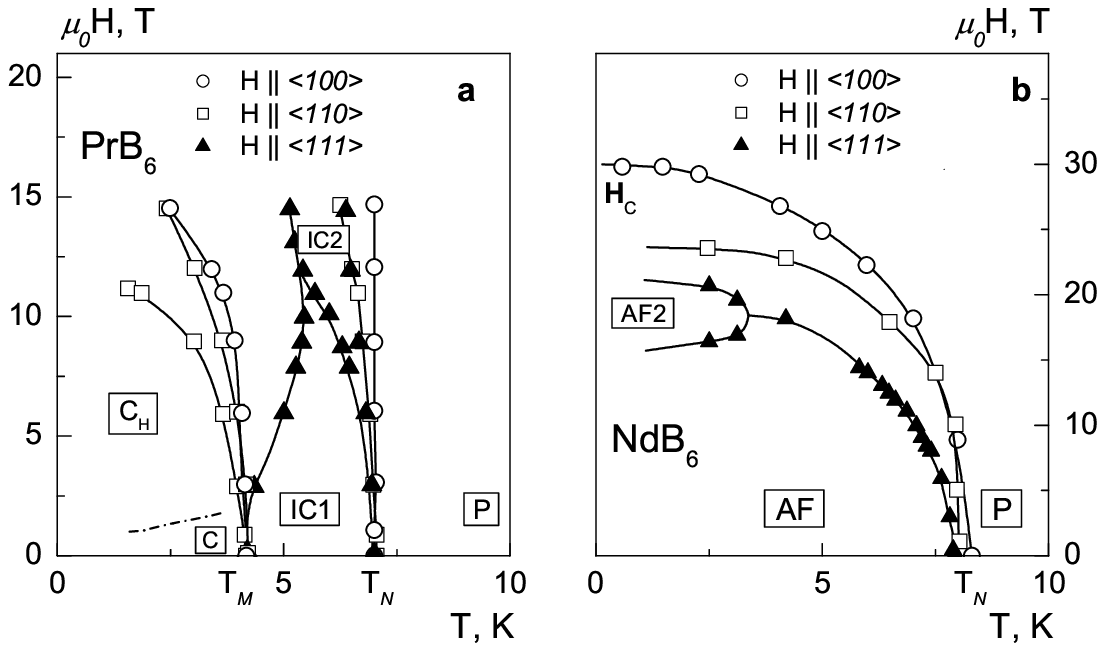}
   \caption{H-T magnetic phase diagram of (a) PrB$_6$ and (b)
   NdB$_6$ for various directions of magnetic fields \textbf{H}$||\langle100\rangle$,
   \textbf{H}$||\langle110\rangle$, \textbf{H}$||\langle111\rangle$.
   The data are taken from \cite{Ser7}-\cite{Iwa9} and \cite{Awa10}-\cite{Yon12} for PrB$_6$ and NdB$_6$ correspondingly.
   P, IC, C and C$_H$ denote paramagnetic, incommensurate, commensurate and
   collinear magnetic phases of PrB$_6$. C-C$_H$ phase transition
   is shown in panel (a) by dash-dotted line (see text for details). }\label{FigX1}
\end{figure*}

A promising challenge to shed more light on the complicated
interplay between electronic and magnetic degrees of freedom in
PrB$_6$ and NdB$_6$ is provided by the study of Hall effect, which
is known to be sensitive to the variation of the FS volume and
topology. Available information about Hall coefficient behaviour
in PrB$_6$ and NdB$_6$ given in \cite{Onu13, Ser16, Stan17} is
fragmentary and controversial. In particular, the measurements of
Hall resistivity performed in low ($\mu_0$H$\approx0.8$T for
PrB$_6$ and NdB$_6$ \cite{Onu13}) and moderate (up to $15$T for
NdB$_6$ \cite{Ser16}) magnetic fields showed that the Hall
coefficient of these compounds doesn't depend on temperature in
paramagnetic state T$>$T$_N$. This observation contradicts
evidently to the large variation of Hall resistivity (more than by
a factor of $2$) established in low magnetic fields
$\mu_0$H$=0.1$T and attributed to anomalous Hall effect in
paramagnetic state of NdB$_6$ \cite{Stan17}. The discrepancy in
experimental results and a lack of Hall effect data for strong
enough magnetic fields makes it difficult to explain correctly
both the exchange parameters' evolution and the magnetic phase
diagrams observed in the compounds of RB$_6$ family.

\section*{2. EXPERIMENTAL DETAILS}

This article reports on the study of Hall effect carried out on
PrB$_6$ and NdB$_6$ single crystals at temperatures
$2$K$<$T$<300$K in magnetic fields $\mu_0$H$\leq8$T. The single
crystals of rare earth hexaborides RB$_6$ (R$=$Pr, Nd) were grown
by crucible-free inductive zone melting. X-ray diffraction and
electron microprobe analysis were used to control the high quality
of the grown crystals. The rectangular bar samples cut from the
single crystal rods were etched in diluted nitric acid to
eliminate the surface defects induced by mechanical treatment.

The angular dependencies of Hall resistivity $\rho_H(\varphi)$
have been measured by the stepwise sample rotation technique in
fixed magnetic field perpendicular to rotation axis \cite{Sl3}. In these experiments the
$\rho_H(\varphi)$ data are produced by the variation of the angle
between the normal to the plane of the sample \textbf{n} and
magnetic field \textbf{H} as a result of change in the scalar
product (\textbf{n}, \textbf{H}), which in turn modulates the Hall
signal by harmonic law. Note that the peak-to-peak value deduced
from the $\rho_H(\varphi)$ studies as the difference
$\rho_H$(+H)$-\rho_H$($-$H) equals to this one extracted in the
commonly used field sweeping technique of Hall resistivity
measurements. The \emph{dc}-current was applied along
$\langle110\rangle$ axis taken to be parallel to the axis of
rotation. High stability of magnetic field ($\Delta
$H/H$\sim10^{-5}$ at $\mu_0$H$=8$T) and temperature ($\Delta
$T$\sim0.01$K) required for this high precision measurements was
achieved with the help of Cryotel SMPS$-60$ superconducting magnet
power supply and Cryotel TC $1.5/300$ temperature controller
operating with LakeShore CX$-1050$ temperature sensor.

\section*{3. EXPERIMENTAL RESULTS}
\subsection*{\emph{3.1. Temperature behavior of resistivity.}}

\begin{figure}
\includegraphics{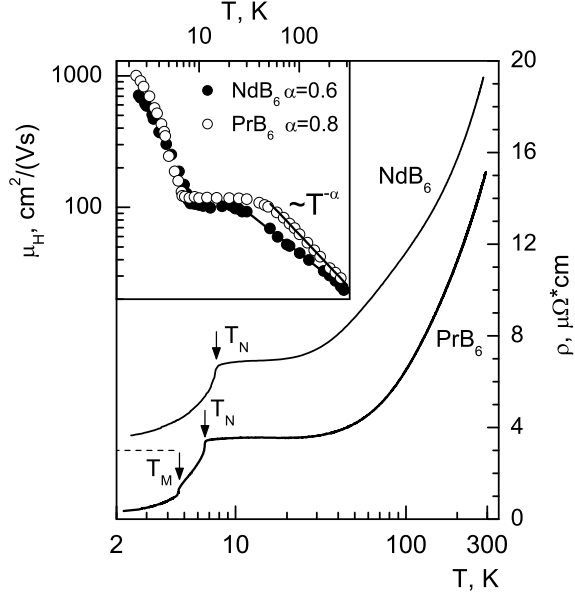}
   \caption{Temperature behaviour of resistivity $\rho$(T)
   measured for PrB$_6$ and NdB$_6$ in zero magnetic field.
   The $\rho$(T) curve for NdB$_6$ is shifted upwards by $3
   \mu\Omega\cdot$cm
   along vertical axis for convenience. Arrows point
   to magnetic phase transition temperatures.
   Inset shows the behaviour of Hall mobility $\mu_H$(T)=R$_H$(T)/$\rho$(T)
   estimated from experimental data (see text).
   The solid lines in the inset represent
   the power law dependences $\mu_H$(T)$\sim$T$^{-\alpha}$  with the exponents given in the legend.}\label{FigX2}
\end{figure}

The temperature dependences of resistivity $\rho$(T) measured for
PrB$_6$ and NdB$_6$ in zero magnetic field are presented in Fig.2.
It is seen from the data of Fig.2 that in paramagnetic state
(T$>$T$_N$) the $\rho$(T) curves demonstrate temperature behavior
to be typical for metals. Rather high values of residual
resistivity in paramagnetic phases
($\rho$($10$K)$=3.5\mu\Omega\cdot$cm and
$\rho$($10$K)$=3.9\mu\Omega\cdot$cm for PrB$_6$ and NdB$_6$,
respectively) point to strong magnetic scattering of itinerant
electrons. The onset of AF state results in a prominent decrease
of resistivity down to the values of $0.4\mu\Omega\cdot$cm and
$0.7\mu\Omega\cdot$cm measured at T$=2.5$K in PrB$_6$ and NdB$_6$,
respectively. The absolute values of resistivity and the magnetic
phase transition temperatures determined from $\rho$(T)
measurements (T$_N\approx 6.7$K, T$_M\approx4.6$K for PrB$_6$ and
T$_N\approx7.7$K for NdB$_6$, see Fig.2) agree with the previous
data \cite{Kob8, Onu13, Ser16} proving the high quality of the
samples under investigation.

\subsection*{\emph{3.2. Hall effect in PrB$_6$ and NdB$_6$}}
The angular dependencies of the Hall resistivity $\rho_H(\varphi)$
measured for PrB$_6$ and NdB$_6$ are presented in Fig.3a and
Fig.3b, correspondingly. The angular dependences of Hall
resistivity $\rho_H$ observed in the paramagnetic phases of these
compounds could be well described by simple cosine law
$\rho_H(\varphi)$= $\rho_{H0}+ \rho_{H1}cos\varphi$  (Fig.3,
T$\geq8$K). In vicinity of AF phase transition (T$\sim$T$_N$)
$\rho_H(\varphi)$ curves deviate from the simple cosine behaviour
and additional contribution to the Hall effect from second
harmonic $\rho_H(\varphi)\sim\cos2\varphi$ appears in the
experimental data (see, e.g., the curves T$= 5.9$K in Fig.3a for
PrB$_6$ and T$=6.5$K and $4.1$K in Fig.3b for NdB$_6$). As a
result, the $\rho_H(\varphi)$ dependences may be fitted by
relation
\begin{equation}\label{Eq.1}
   \rho_H(\varphi)= \rho_{H0}+ \rho_{H1}cos\left(\varphi\right)+
\rho_{H2}cos\left(2\varphi - \Delta\varphi\right),
\end{equation}
where  $\rho_{H0}$ is a constant value arising due to the
misalignment of Hall probes,  $\rho_{H1}$ and $\rho_{H2}$ are the
amplitudes of the main and second harmonic and $\Delta\varphi$ is
the phase shift of the second harmonic.

\begin{figure*}
\includegraphics[scale=1.2]{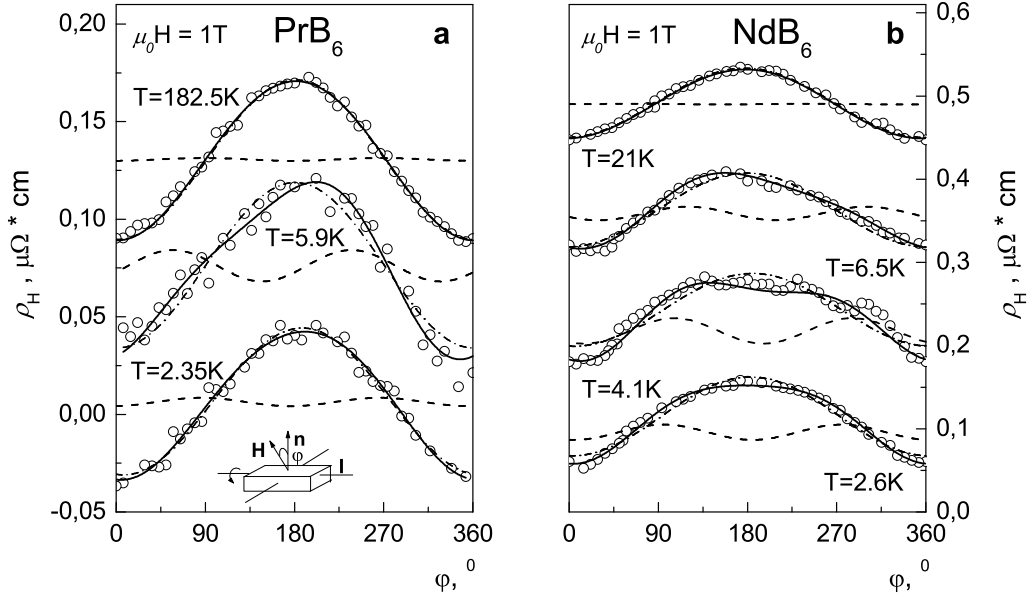}
   \caption{Angular dependencies of Hall resistivity $\rho_H(\varphi,$ T$_0$)
   measured in magnetic field $\mu_0$H$=1$T for (a) PrB$_6$ and (b) NdB$_6$.
   The curves are shifted along a vertical axis for convenience.
   The lines correspond to the contributions of the first (dash-dot)
   and second (dash) harmonics as well as the sum of the harmonics
   (solid) (see Eq.1). The inset in panel (a) illustrates
   the experimental scheme with the sample rotation.}\label{FigX3}
\end{figure*}

Note that the transverse configuration of the Hall experiment used
in our study (the rotational axis of the sample is parallel to
vector \textbf{I} and perpendicular to vector \textbf{H}; see the
inset in Fig.3a) allows to minimize the spurious magnetoresistance
contribution to the Hall signal, which could also result in even
harmonic \cite{Sl3, Sl18}. To estimate independently the possible
contribution to Hall effect induced by anisotropic
magnetoresistance $\sim\cos(2\varphi)$, the angular dependences of
Hall resistance and magnetoresistance were recorded simultaneously
with subsequent scaling of magnetoresistance data $\rho(\varphi)$
to $\rho_{H0}$ emerging due to misalignment of Hall probes. Our
estimations lead to conclusion that the magnetoresistive component
has no appreciable effect on Hall resistivity $\rho_H(\varphi)$ in
magnetic fields up to $8$T.

The approach described above was earlier applied to separate the
various contributions to Hall effect in the AF phases of
antiferromagnet CeAl$_2$ \cite{Sl18} and heavy fermion compound
CeB$_6$ \cite{Sl3}. Besides, similar technique was successfully
used to establish and explain the complicated behavior of Hall
coefficient in rare earth dodecaborides RB$_{12}$ \cite{Sl19,
Sl20} and in metallic systems with heavy fermions CeAl$_3$
\cite{Sl21} and quantum critical behavior CeCu$_{6-x}$Au$_x$
\cite{Sl22}.

In present study the $\rho_H(\varphi)$ data were fitted by (Eq.1)
both in paramagnetic and AF phases of RB$_6$ (R$=$Pr, Nd). As a
result, the temperature and magnetic field behaviour of the Hall
resistivity $\rho_{H1}$(T,H), even harmonic term $\rho_{H2}$(T,H)
and the phase shift  $\Delta\varphi$(T,H) to be deduced from the
experimental data are presented and discussed in the next
sections.

\subsection*{\emph{3.3. Temperature and magnetic field dependences of Hall coefficient in PrB$_6$ and NdB$_6$}}
The amplitude of the first harmonic term  $\rho_{H1}$ was used to
calculate the Hall coefficient R$_H$(T)=$\rho_{H1}$(T)/H
(d is the sample thickness). The temperature dependencies of
R$_H$(T) obtained for PrB$_6$ and NdB$_6$ are presented in Fig.4a
and Fig.4b, correspondingly. The R$_H$(T) data for nonmagnetic
reference compound LaB$_6$ ($4f^0$ configuration) is also shown in
Fig.4a for comparison. It is found that Hall coefficient increases
for both PrB$_6$ and LaB$_6$ above liquid nitrogen temperature
(Fig.4a), but for NdB$_6$ it decreases only slightly when
temperature rises in the interval $70$K$\leq$T$\leq300$K
(Fig.4b). At the same time, the data in Fig.4 demonstrate that
Hall coefficient measured in magnetic field $\mu_0$H$=1$T for
PrB$_6$ and NdB$_6$ doesn't depend noticeably on temperature in
the range of $8\div70$K. Note also that in this temperature
interval the estimated values of Hall coefficient
R$_H$(PrB$_6)\approx-(4.2\pm0.1)\cdot10^{-4}$ cm$^3$/C and
R$_H$(NdB$_6)\approx-(4.1\pm0.1)\cdot10^{-4}$ cm$^3$/C agree with
the data of Onuki et al \cite{Onu13} (see also Fig.5a).

\begin{figure*}
\includegraphics[scale=1.2]{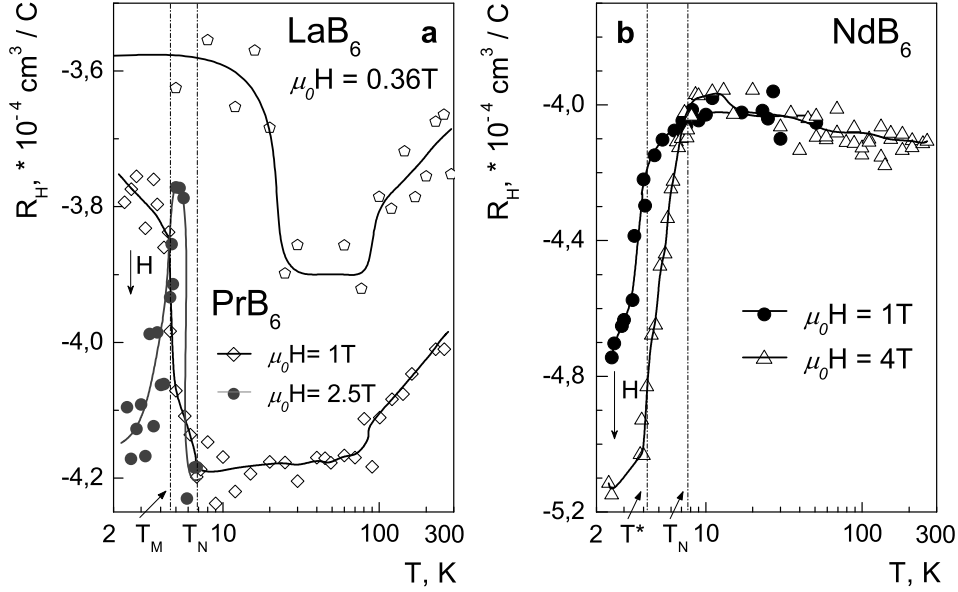}
   \caption{Temperature dependences of Hall coefficient
    R$_H$(T) for (a) LaB$_6$, PrB$_6$ and (b) NdB$_6$
    in magnetic fields $\mu_0$H$\leq4$T. Solid lines are drawn to guide for eye.}\label{FigX4}
\end{figure*}

The temperature dependence of the R$_H$(T, $\mu_0$H$=1$T) in the
paramagnetic phase of NdB$_6$ (Fig.4b) contradicts to the low
field Hall effect data obtained in \cite{Stan17}. In contrast to
the present work, Hall resistivity measured for the same current
direction \textbf{I}$||\langle110\rangle$ in \cite{Stan17}
demonstrates a pronounced peak of $\rho_H$(T) at T$_N\sim8$K,
which is followed by a gradual diminishing of R$_H$ when the
temperature increases up to room value (Fig.5a). To verify this
issue the magnetic field dependences of Hall resistivity
$\rho_H$(H) were measured for NdB$_6$ at temperatures below and
above T$_N$ in magnetic fields up to $8$T. Special attention was
paid to the range of low magnetic fields $\mu_0$H$\leq1$T, where
drastic difference of R$_H$(T) values measured in \cite{Onu13,
Stan17} is observed (to compare R$_H$(10
K)$=-(4.6\pm0.9)\cdot10^{-4}$ cm$^3$/C \cite{Onu13} and
R$_H$(10 K)$=$$\rho_H$(10 K)$-4\cdot 10^{-4}$ cm$^3$/C
\cite{Stan17}, see Fig.5a). The comparison between the $\rho_H$(H)
data obtained in the present study and in \cite{Stan17} (Fig.5b)
shows that Hall resistivity depends linearly on magnetic field
demonstrating no contribution of second harmonic in the
considered range of temperatures and magnetic fields. As a result,
the data of Fig.5 allow to conclude that in the paramagnetic state
of NdB$_6$ Hall coefficient doesn't depend noticeably on
temperature and magnetic field showing only small (less than
$10\%$) variation of R$_H$ (Fig.5a). Therefore our data do not
confirm earlier observed Hall coefficient anomaly at T$\sim$T$_N$
\cite{Stan17} (Fig.5a) and it makes questionable also the
interpretation in terms of the anomalous Hall effect in the
paramagnetic phase of NdB$_6$ proposed in \cite{Stan17}.

The transition to AF state is accompanied by a different variation
of the low field Hall coefficient in PrB$_6$ and NdB$_6$. For
T$<$T$_N$ the absolute value of R$_H$(T) in PrB$_6$ decreases
drastically ($\Delta$R$_H$/R$_H\sim10\%$) when temperature is
lowered (see curve $\mu_0$H$=1$T in Fig.4a). At the same time, the
absolute value of R$_H$(T) in NdB$_6$ increases noticeably
($\sim15\%$ for $\mu_0$H$=1$T) just below the Neel temperature
(Fig.4b). In our opinion, this difference in R$_H$(T) behaviour
may be attributed to the peculiarities of AF phases and it will be
discussed in the last section.

\subsection*{\emph{3.4. Second harmonic contribution in Hall effect}}
In paramagnetic phase of PrB$_6$ and NdB$_6$ the contribution of
$\rho_{H2}$ is very small as compared to the first harmonic.
However, the amplitude  $\rho_{H2}$ rises drastically when AF
state sets up in PrB$_6$ and NdB$_6$ (Fig.6). It is worth noting
that for PrB$_6$ the term $\rho_{H2}$ contributes essentially to
Hall signal in low fields $\mu_0$H$\leq1$T only in the
incommensurate AF phase and the component was evidently observed
in the range T$_M$$<$T$<$T$_N$ (see Fig.1a), while for NdB$_6$ the
temperature dependence of $\rho_{H2}$(T) is characterized by a
pronounced maximum at intermediate temperature T$^{\ast}\sim4$K,
which is well below T$_N$ (Fig.6). The elevation of magnetic field
results in an essentially different behaviour of the second
harmonic term in PrB$_6$ and NdB$_6$. Indeed, a pronounced
increase of the $\rho_{H2}$(T) values in PrB$_6$ is accompanied by
broadening of the peak in magnetic field (Fig.6a). On the
contrary, for NdB$_6$ the $\rho_{H2}$(T) maximum value decreases
evidently for $\mu_0$H$>1$T (see, e.g., data for $\mu_0$H$=4$T in
Fig.6b).

Interesting that noticeable anomalies of both magnetoresistance
\cite{An23, An24} and C$_{44}$ elastic constant \cite{Nak25}
temperature dependencies have been earlier detected in the
vicinity of temperature T$^{\ast}$ for NdB$_6$. These features at
T$^{\ast}\sim4$K may indicate on the possible changes of
electronic and/or magnetic structure occurred in the commensurate
AF phase of NdB$_6$ well below the Neel temperature. However, a
detailed investigation of magnetic and charge transport parameters
of NdB$_6$ need to be carried out to shed more light on the origin
of the Hall effect anomalies observed at T$^{\ast}$ in AF phase of
NdB$_6$.

\begin{figure*}
\includegraphics[scale=1.2]{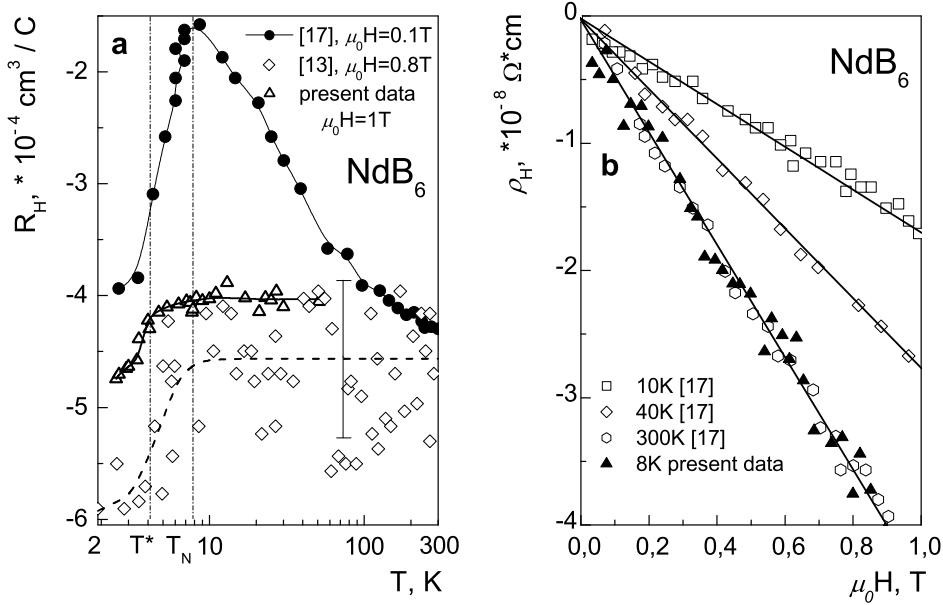}
   \caption{(a) The comparison of R$_H$(T) temperature dependences obtained
   for NdB$_6$ in this study with those ones reported in \cite{Onu13} and \cite{Stan17}.
   Panel (b) presents the field dependences of low field ($\mu_0$H$\leq1$T)
   Hall resistivity $\rho_H$(H) in comparison with the data of \cite{Stan17}.}\label{FigX5}
\end{figure*}

\section*{4. DISCUSSION}
\subsection*{\emph{4.1. Paramagnetic phase of PrB$_6$ and NdB$_6$ (T$>$T$_N$)}}
High accuracy of data obtained allows us to estimate the number of
conduction electrons per unit cell in the paramagnetic phase of
rare earth hexaborides under investigation. The values
n/n$_{4f}$(PrB$_6)\approx1.05\pm0.01$ and
n/n$_{4f}$(NdB$_6)\approx1.09\pm0.01$, which are established just
above the Neel temperature (at $10$K) in these magnetic
hexaborides, are comparable with charge carriers' concentration
previously estimated for CeB$_6$ (n/n$_{4f}\approx1.00$
\cite{Sl3}). An approximately linear increase of n/n$_{4f}$
parameter vs $4f$-shell occupation in the magnetic hexaborides
agrees well with the results obtained in \cite{Onu26, Beh27},
where a remarkable expansion of the small FS electron pockets was
detected in the light RB$_6$ compounds. The increase of n/n$_{4f}$
produced by FS changes is a factor, which is responsible for a
variation of both RKKY-function $\Sigma$F($2$k$_F$R$_i$) (here
k$_F$ and R$_i$ denote Fermi wavevector and the distance between
the magnetic moments of the rare earth ions, respectively) and
indirect exchange interaction. According to de Gennes approach
\cite{Gen28}, the Neel temperature T$_N$, reduced concentration
n/n$_{4f}$, exchange constant J$_{ex}$, Fermi energy E$_F$, de
Gennes factor G$=$(g$-1$)$^2$J(J+1) and RKKY-function
$\Sigma$F($2$k$_F$R$_i$) are related by the expression

\begin{equation}\label{Eq.2}
    T_N=\frac{3\pi}{4}\left(\frac{n}{n_{4f}}\right)^2G\left(\frac{J_{ex}^2}{E_F}\right)\Sigma F(2k_FR_i).
\end{equation}

Taking into account both the strong variation of the de Gennes
factor (G(PrB$_6)\approx0.8$ and G(NdB$_6)\approx1.84$) and $5\%$
increase in the charge carriers concentration n/n$_{4f}$ together
with only small changes in the lattice constant and, hence, in the
J$_{ex}$ exchange parameter, one needs to propose an essential
decrease of the $\Sigma$F($2$k$_F$R$_i$) values from PrB$_6$ to
NdB$_6$, which is necessary to explain relatively small changes in
Neel temperatures from T$_N$(PrB$_6)\approx6.7$K to
T$_N$(NdB$_6)\approx7.7$K. Indeed, the pronounced lowering of
RKKY-function was predicted for RB$_6$ compounds when the
n/n$_{4f}$ ratio increases in the range $1-1.2$ \cite{Bar29}, but
more detailed calculations are necessary to estimate
quantitatively the variation of T$_N$ in the rare earth
hexaborides family.

\subsection*{\emph{4.2. Antiferromagnetic state of PrB$_6$ and NdB$_6$ (T$<$T$_N$) }}
\begin{figure*}
\includegraphics[scale=1.2]{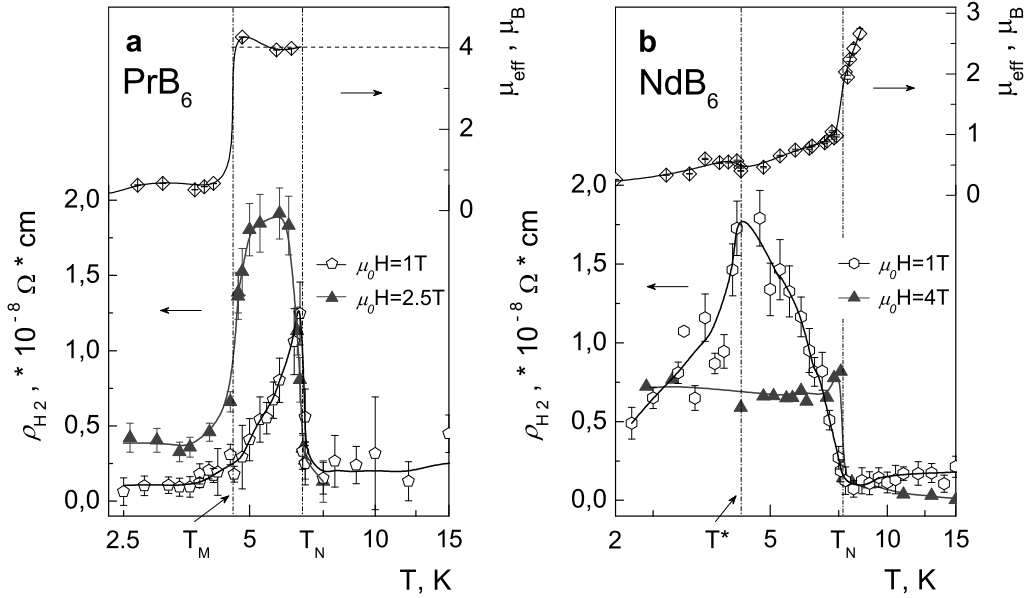}
   \caption{Temperature dependencies of the amplitude of
   the second harmonic term $\rho_{H2}$ measured for (a) PrB$_6$ and (b)
   NdB$_6$
   in magnetic fields $\mu_0$H$\leq4$T. The upper curves on the both panels
   represent the temperature dependencies of the effective
   magnetic moment $\mu_{eff}$ \cite{An24}.}\label{FigX6}
\end{figure*}

When discussing the complicated behavior of Hall effect in AF
phases of PrB$_6$ and NdB$_6$ (see sections $3.3-3.4$) it is worth
to mention that the magnetic structures of PrB$_6$ and NdB$_6$
have different parameters in low magnetic field region. The
magnetic unit cell of PrB$_6$ C-phase involves $32$ structural
unit cells \cite{Mc30} and characterizes by a wave vector
\textbf{Q}$=(1/4,1/4,1/2)$ while simple doubling of the structural
unit cell is observed in NdB$_6$ \cite{Mc31}. On the contrary, the
structure of PrB$_6$ commensurate C$_H$ magnetic phase formed in
moderate magnetic fields (Fig.1a) is similar to that one of
commensurate AF phase observed in NdB$_6$ in magnetic fields below
$15$T \cite{Ser7, Iwa9, Awa10}. In this respect the dramatic
change of Hall coefficient R$_H$ found in PrB$_6$ with the
increase of magnetic field from $1$T to $2.5$T (Fig.4a) may be
definitely associated with the crossing of the C-C$_H$ phase
boundary below T$_M$.

To obtain more information about the variation of Hall coefficient
when entering from C to C$_H$ phases of PrB$_6$, the magnetic
field dependencies of R$_H$(H,T$_0$) have been obtained from the
experimental data for fixed temperatures T$_0<$T$_M$ in the range
$\mu_0$H$\leq6$T. The R$_H$(H,T$_0$) data shown in Fig.7a allow to
detect clearly the C-C$_H$ phase transition in PrB$_6$
establishing a positive slope of C-C$_H$ phase boundary in
agreement with the results \cite{Kob8, Iwa9} (see also Fig.1a).
So, the similar behaviour of the Hall coefficient temperature
dependencies for PrB$_6$ at T$<$T$_M$ (see data for
$\mu_0$H$=2.5$T in Fig.5a) and for NdB$_6$ at T$<$T$_N$ (Fig.5b)
may be likely understood assuming identical magnetic structures
(simple type I antiferromagnet with ordering vector \textbf{Q}$=
(0,0,1/2)$ ) developing in the AF phase of NdB$_6$ (Fig.1b) and in
the C$_H$-phase of PrB$_6$ (Fig.1a).

The strong renormalization of the Hall coefficient R$_H$(T) in the
commensurate phases of the studied antiferromagnets (Figs.5,7) may
be attributed to the reconstruction of FS below the AF phase
transition. The electronic structure of the AF state in these
hexaborides may be properly understood through folding of the
paramagnetic band structure. Following to the FS reconstruction of
\cite{Kub15}, the simple cubic Brillouin zone in the paramagnetic
state of RB$_6$ is reduced by this folding procedure to the
tetragonal one in the AF phase of NdB$_6$. As a result, two kinds
of electron sheets and one hole sheet appear in the AF phase of
NdB$_6$ causing to the remarkable changes in the Hall coefficient
behaviour with temperature both in weak and strong magnetic field
regimes. However, the independent verification of this scenario
requires a comparative study of Hall effect and magnetization to
be carried out in strong enough magnetic fields.

\begin{figure*}
\includegraphics[scale=1.15]{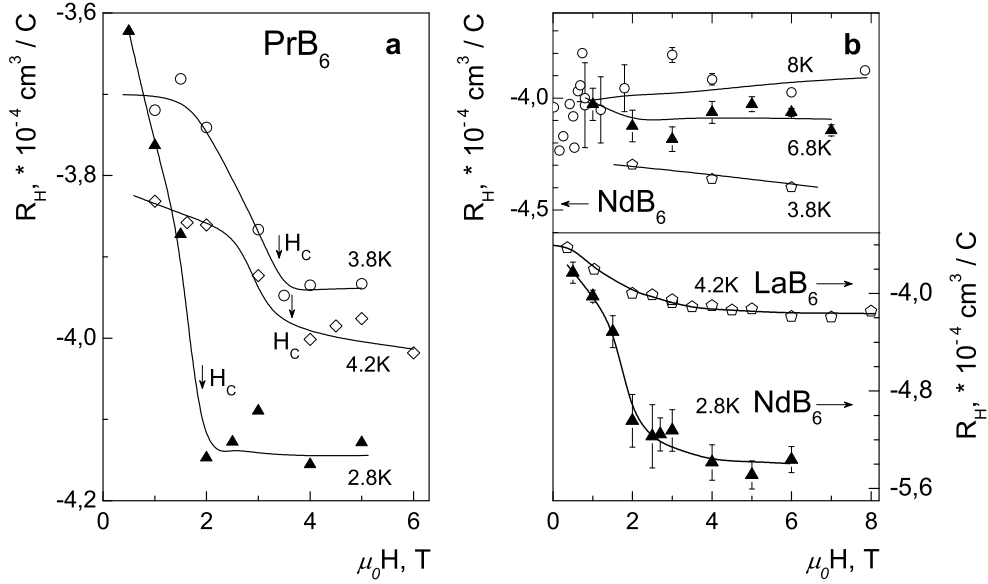}
   \caption{ The field dependencies of Hall coefficient R$_H$(H)
   for (a) PrB$_6$ and (b) NdB$_6$ in the temperature region T$\leq8$K.
   Arrows in panel (a) point to the characteristic fields of C-C$_H$ transition.
   The data of LaB$_6$ are given for comparison in panel (b).}\label{FigX7}
\end{figure*}

The resistivity (Fig.2) and Hall effect (Fig.4) data were used to
estimate Hall mobility $\mu_H$(T)$=|$R$_H$(T)$|/\rho$(T) for
PrB$_6$ and NdB$_6$ (inset in Fig.2). In commensurate AF phases of
these compounds Hall mobility gets values
$\mu_H$(PrB$_6)\approx1000$ cm$^2$/(V$\cdot$s) and
$\mu_H$(NdB$_6)\approx710$ cm$^2$/(V$\cdot$s). In paramagnetic
state in the range $45$K$\leq$T$\leq300$K the $\mu_H$(T) curves
could be well fitted by the power law dependence
$\mu_H\sim$T$^{-\alpha}$ with the exponents
$\alpha$(PrB$_6)\sim0.8$ and $\alpha$(NdB$_6)\sim0.6$ (inset in
Fig.2). The observed decrease of both Hall mobility $\mu_H$ and
exponent $\alpha$ when moving from PrB$_6$ to NdB$_6$ may be
associated with the enhancement of magnetic scattering of
itinerant electrons on the localized magnetic moments of
$4f$-states of R$^{3+}$ ions (R=Pr, Nd) that agrees well with de
Gennes scaling rule \cite{Gen28}.

Finally, it is worth to note a very large difference in the low
temperature values of Hall mobility $\mu_H$ estimated from the
experimental data for non-magnetic reference compound LaB$_6$
($\mu_H(4.2$K)$\approx18000$ cm$^2$/(V$\cdot$s),
$\rho(4.2$K)$\approx0.016$ $\mu\Omega\cdot$cm) and
antiferromagnetic PrB$_6$ and NdB$_6$
($\mu_H(4.2$K)$\approx370-430$ cm$^2$/(V$\cdot$s), see inset in
Fig.2). As a result, the remarkable changes of the Hall
coefficient in magnetic field observed in the present study for
LaB$_6$ and NdB$_6$ (Fig. 7b) should be explained by quite
different factors. In the case of nonmagnetic LaB$_6$ the strong
enough variation of R$_H$(H, $4.2$K) (by $\sim16\%$) could be
certainly attributed to transition from weak ($\omega\tau\ll1$,
where and are cyclotron frequency and charge carriers' relaxation
time, correspondingly) to strong ($\omega\tau\gg1$) magnetic field
regime \cite{Lif32}. However, rather low mobility of charge
carriers in NdB$_6$ doesn't allow to interpret the drastic changes
of R$_H$(H) observed in the AF phase of NdB$_6$ (see, e.g., curve
for T$=2.8$K in Fig.7b) in terms of the $\omega\tau$ approach.

To explain the anomalous behaviour of Hall coefficient in the
studied rare earth hexaborides it is necessary to address to the
results of transverse magnetoresistance study performed recently
for these compounds \cite{An24}. In particular, it was shown
\cite{An24} that nanoscale magnetic clusters with strongly
renormalized effective magnetic moments
$\mu_{eff}$(PrB$_6)\approx4\mu_B$ and
$\mu_{eff}$(NdB$_6)\approx2.5\mu_B$ are formed just above the Neel
temperature in these compounds (see also $\mu_{eff}$(T) curves in
Fig.6). The unit cell magnetic clusters' formation attributed to
the exchange induced $5d$-states' spin polarization \cite{An24}
was proposed to be responsible both for the effects of
density-of-states renormalization and the formation of additional
$5d$-component in the magnetic structure of these unusual
antiferromagnets. In such a case the interaction between the spin
polarized component of $5d$-states and the magnetic structure of
$4f$ magnetic moments in the rare earth hexaborides under
investigation could be considered as an important reason resulting
both in the appearance of the magnetic anisotropy in AF-phase
\cite{An23, An24} and in the noticeable renormalization of Hall
effect observed just below T$_N$ in RB$_6$ under investigation
(Fig.4).

\section*{5. CONCLUSION}
To summarize, the Hall effect in the antiferromagnetic metals
PrB$_6$ and NdB$_6$ has been studied at temperatures in the range
$2-300$K in magnetic fields up to $8$T. The detailed comparison
between the R$_H$(T) temperature dependencies obtained in present
investigation and those ones reported earlier in \cite{Onu13,
Stan17} allows establishing temperature independent behaviour of
R$_H$(T) in paramagnetic state of these two RB$_6$ compounds
excluding the interpretation \cite{Stan17} in terms of anomalous
paramagnetic Hall effect in NdB$_6$. Within the analysis based on
the de Gennes approach an essential decrease of RKKY-function
amplitude from PrB$_6$ to NdB$_6$ is suggested to be the main
reason of the close magnitudes of Neel temperatures in these
magnetic hexaborides with very different values of the de Gennes
factors. Additionally, quite different behaviour of R$_H$(T) was
found below T$_N$ for PrB$_6$ and NdB$_6$ in low magnetic fields.
It was shown that the transition to the commensurate C$_H$ phase
in PrB$_6$ is accompanied by the pronounced (up to $10\%$)
decrease of Hall coefficient R$_H$ at liquid helium temperatures.
As a result the temperature behaviour of R$_H$(T) in the C$_H$
phase of PrB$_6$ is proved to be similar to that one found in the
commensurate AF phase of NdB$_6$. Our findings favour the
enhancement of the magnetic scattering of the charge carriers on
the localized magnetic moments of R$^{3+}$ ions when moving from
PrB$_6$ to NdB$_6$. The observed variation of Hall coefficient in
the AF phases of these RB$_6$ compounds is supposed to be induced
by the effects of paramagnetic FS structure folding and the
effects of density-of-states renormalization, which could be
attributed to the magnetic polarization of $5d$-states both in AF
and paramagnetic states of the hexaborides.

\section*{6. ACKNOWLEDGEMENTS}
Support by the RAS Program "Strongly Correlated Electrons in
semiconductors, metals, superconductors and magnetic materials"
and RFBR project $10-02-00998$ is acknowledged. Helpful
discussions with Prof. G.E. Grechnev are greatly appreciated.

{}
\end{document}